# Biosensing using Functionally Graded Piezoelectric MEMS Resonators

Meysam T. Chorsi[1]

[1] Department of Mechanical Engineering, University of Connecticut, Storrs, CT 06269 USA

**Abstract**

Nonlinear dynamics of a two-side electro-statically actuated capacitive micro-beam is studied. The piezoelectric actuation leads to the generation of an axial force along the length of the micro-beam and this is used as a tuning tool to shift the primary resonance of the micro-resonator. The governing equation of motion is derived by minimization of the Hamiltonian and generalized to the viscously damped systems. The periodic solutions in the vicinity of the primary resonance are detected and their stability is investigated. The basins of attraction conforming to three individual periodic orbits are determined. The outcomes show that the higher the amplitude of the periodic orbit, the smaller is the area of the attractor.

1. Introduction

Today the application of MEMS has considerably increased and therefore study of their behavior is of great importance [1-4]. A group of papers in the literature of the dynamics of capacitive MEM structures are devoted to the so-called Pull-in instability due to a DC voltage [5-7]. Another field of research which has gained a great deal of attention is on the non-linear dynamics of capacitive micro-structures subject to a combination of a bias DC and a harmonic excitation. This type of actuation enables designing low driving voltage and switching time (Radio Frequency) RF switches [8]. [9] studied on the nonlinear response of resonant micro-beam subject to an electric actuation; using a single mode approximation they performed perturbation method to capture the amplitudes of the periodic solutions and studied the stability of the periodic orbits. [10] studied on the non-linear dynamics of the same model proposed by [8], they proposed a novel approach to generate reduced-order models; they reported that a Taylor series expansion of the nonlinear electrostatic term fails to represent the electro-static force especially in the vicinity of the pull-in instability. [11] studied on the same model as proposed by [10, 12], [8, 13, 14] and, they simulated the dynamic behavior and compared the results with numerical solutions and available experimental results in the literature. [15, 16] published papers on the sub and super harmonic resonances of a clamped-clamped capacitive micro-beam subject to the same actuation voltage as reported in [8, 10, 13, 14, 16, 17]. [14] in another paper investigated the hardening/softening behavior in the vicinity of primary resonance of fully clamped capacitive micro-beam. [17] represented analysis of a capacitive clamped-clamped micro-beam under secondary resonance excitation. [18-20] determined the influence of Casimir force on the nonlinear behavior of an electro-statically actuated nano-switch. In the present study a two side electro-statically actuated fully-clamped FGP/silicon micro-beam resonator is investigated. The dynamics of micro beams subject to two side electrostatic actuation is more complicated than the single electro static actuation due to the appearance of chaotic dynamics which is focused on especially in micro mixers [21, 22]. Furthermore, the pull-in threshold due to bias DC voltage occurs in higher voltages in compare with the single electrostatic actuation. The tunability of the pioneer resonant sensors is due to the DC voltage of single electrostatic actuation, however this only make the device tunable only in backward direction. The proposed model is a tunable (both in forward and backward directions) resonant sensor. The innovation of the model is its tunability of resonance frequency and its robustness against pull-in instability due to the two side simultaneous electrostatic actuation. The periodic orbits are captured using the so called shooting method and their stability is investigated. The basins of attraction for some periodic orbits are determined.

## 2. Modeling

Functionally Graded Piezoelectric fully clamped capacitive micro-beam of length $l$, thickness $h$, and width $a$ is proposed. The resonator is subject to two stationary electrodes located on either sides of the micro-beam. Through the upper electrode a combination of DC and a harmonic voltage with amplitude $V_{AC}$ and frequency $\Omega$ is applied. The lower electrode imposes a pure DC voltage the same as that of upper electrode [22].

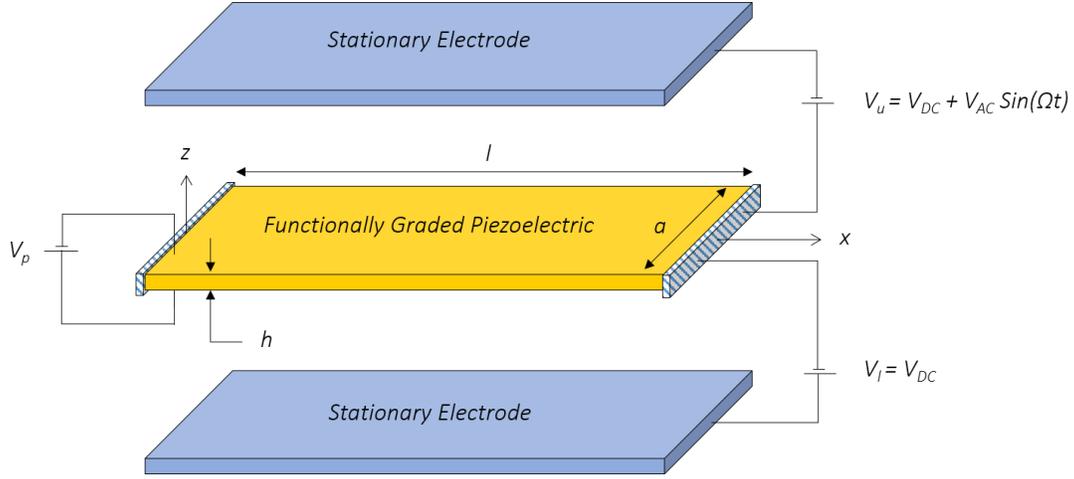

Figure 1: Schematics of the functionally graded piezoelectric micro-resonator (FGPM) and the applied voltages

The mechanical properties associated with silicon and PZT are denoted by subscriptions '$S$' and '$P$' respectively. The mechanical properties vary with respect to power law distribution as:

$$E(z) = E_q e^{\gamma |z|} \tag{1}$$
$$\rho(z) = \rho_q e^{\alpha |z|}$$
$$e_{31}(z) = e_{(31_P)} e^{\mu |z|} - \beta$$

where $E_q, \gamma, \rho_q, \alpha, \mu$ and $\beta$ are constants which are determined based on the following conditions:

$$z = 0: \quad MP = MP_0 = P_{S_0} MP_S + P_{P_0} MP_P \tag{2}$$
$$z = \frac{h}{2}: \quad MP = MP_u = P_{S_u} MP_S + P_{P_u} MP_P$$

Considering the conditions in Eq. (2),Eq. (1) reduces to :

$$E(z) = (E_0) e^{\frac{2}{h}|z|\ln(\frac{E_u}{E_0})} \tag{3}$$
$$\rho(z) = (\rho_0) e^{\frac{2}{h}|z|\ln(\frac{\rho_u}{\rho_0})}$$
$$e_{31}(z) = e_{31_P}(e^{\frac{2}{h}|z|\ln(1 - P_{P_0} + P_{P_u})} - 1 + P_{P_0})$$

The strain energy due to mechanical bending is expressed as:

$$U_b = \int \frac{\varepsilon_b \sigma_b}{2} dv \qquad (4)$$

$$= \int \frac{E}{2}(-z\frac{\partial^2 w}{\partial x^2})(-z\frac{\partial^2 w}{\partial x^2})dv$$

where $\varepsilon_b$ and $\sigma_b$ are the strain and stress fields due to the bending and $dv$ is the volume of infinite small element; considering the geometry of the micro-beam and simplifying, equation (4), reduces to:

$$U_b = \frac{(EI)_{eq}}{2}\int (\frac{\partial^2 w}{\partial x^2})^2 dx \qquad (5)$$

where :

$$(EI_{yy})_{eq} = \int_{-\frac{h}{2}}^{\frac{h}{2}} aE(z)z^2 dz \qquad (6)$$

The clamped-clamped boundary condition imposes the extended length of the beam ($l'$) become more than the initial length $l$ which leads to the introduction of an axial stress and accordingly an axial force denoted as:

$$F_a = \frac{(EA)_{eq}}{l}(l'-l) \approx \frac{(EA)_{eq}}{2l}\int_0^l \left(\frac{\partial w}{\partial x}\right)^2 dx \qquad (7)$$

where :

$$(EA)_{eq} = \int_{-\frac{h}{2}}^{\frac{h}{2}} aE(z)dz \qquad (8)$$

Considering $L \gg w$, the stretched length ($l'$) is obtained by integrating the arc length "$ds$" as [23]:

$$l' = \int_0^l ds \approx \int_0^l \sqrt{1+\left(\frac{\partial w}{\partial x}\right)^2} dx = l + \frac{1}{2}\int_0^l \left(\frac{\partial w}{\partial x}\right)^2 dx \qquad (9)$$

The corresponding strain energy due to the mid-plane stretching is:

$$U_a = \frac{1}{2}F_a(l'-l) \qquad (10)$$

Substituting equations (7) and (9) into equation (10), the strain energy due to the mid-plane stretching reduces to:

$$U_a = \frac{(EA)_{eq}}{8l}\left(\int_0^l \left(\frac{\partial w}{\partial x}\right)^2\right)^2 \tag{11}$$

and considering the direction of the applied electrical field ($E_3$) the axial stress due to the piezoelectric actuation reduces to:

$$\sigma_1 = -e_{31} E_3 \tag{12}$$

where $e_{31}$ is the corresponding piezoelectric voltage constant (Coulomb/m²); Considering $E_3 = V_p/h_p$, the axial force due to the piezoelectric actuation reduces to:

$$F_p = \int_{A_p} \sigma_1 dA_p = \frac{2aV_p}{h} \int_0^{h/2} -e_{31}(z)dz \tag{13}$$

The strain potential energy due to the axial piezoelectric force is:

$$U_p = F_p(l'-l) = \frac{F_p}{2}\int_0^l \left(\frac{\partial w}{\partial x}\right)^2 dx \tag{14}$$

The total potential energy of the system is as follows:

$$U = U_b + U_a + U_p \tag{15}$$

The kinetic energy of the micro-beam is represented as:

$$T = \frac{(\rho a h)_{eq}}{2} \int_{x=0}^{x=l} \left(\frac{\partial w}{\partial t}\right)^2 dx \tag{16}$$

where:

$$(\rho a h)_{eq} = \int_{-\frac{h}{2}}^{\frac{h}{2}} a\rho(z) dz \tag{17}$$

The work of the electrostatic force from zero deflection to $w(x)$ is expressed as:

$$w_{el} = \int_0^l \left(\int_0^w \frac{\varepsilon_0 a V_u^2}{2(g_0-\zeta)^2} d\zeta\right) dx + \int_0^l \left(\int_0^w \frac{-\varepsilon_0 a V_l^2}{2(g_0+\zeta)^2} d\zeta\right) dx = \\ \frac{\varepsilon_0 a V_u^2}{2}\int_0^l \left(\frac{1}{g_0-w}-\frac{1}{g_0}\right)dx + \frac{\varepsilon_0 a V_l^2}{2}\int_0^l \left(\frac{1}{g_0+w}-\frac{1}{g_0}\right)dx \tag{18}$$

The governing partial differential equation of the motion is obtained by the minimization of the Hamiltonian using variational principle as:

$$\delta \int_0^t H dt = \delta \int_0^t (T - U + w_{el}) dt = 0 \tag{19}$$

Introducing equations (15), (16) and (18) into equation (19), the Hamiltonian reduces to:

$$\delta \int_0^t H dt = \tag{20}$$

$$\int_0^t \left\{ \begin{array}{l} -(EI)_{eq} w'' \delta w' \Big|_0^l + (EI)_{eq} w''' \delta w \Big|_0^l - (EI)_{eq} \int_0^l w^{IV} \delta w dx \\ -F_p w' \delta w \Big|_0^l + F_p \int_0^l w'' \delta w dx - \frac{(EA)_{eq}}{2} \int_0^l w'^2 dx w' \delta w \Big|_0^l + \\ \frac{(EA)_{eq}}{2} \int_0^l w'^2 dx \int_0^l w'' \delta w dx + \frac{\varepsilon_0 a V_u^2}{2} \int_0^l \frac{\delta w}{(g_0 - w)^2} dx - \\ \frac{\varepsilon_0 a V_l^2}{2} \int_0^l \frac{\delta w}{(g_0 + w)^2} dx \end{array} \right\} dt$$

$$+ \int_0^t \left\{ (\rho A)_{eq} \int_0^l \dot{w} \delta w \Big|_0^l - (\rho A)_{eq} \int_0^t \ddot{w} \delta w dt \right\} dx = 0$$

The governing equation and the corresponding boundary conditions reduce to:

$$(EI)_{eq} \frac{\partial^4 w(x,t)}{\partial x^4} + (\rho A)_{eq} \frac{\partial^2 w(x,t)}{\partial t^2} - (F_P + \frac{(EA)_{eq}}{2l} \int_0^l (\frac{\partial w(x,t)}{\partial x})^2 dx) \frac{\partial^2 w(x,t)}{\partial x^2} = \tag{21}$$

$$\frac{\varepsilon_0 a (V_{DC} + V_{AC} \sin(\Omega t))^2}{2(g_0 - w(x,t))^2} - \frac{\varepsilon_0 a V_{DC}^2}{2(g_0 + w(x,t))^2}$$

Subjected to the following boundary conditions:

$$w(0,t) = w(l,t) = 0 \; , \; \frac{\partial w(0,t)}{\partial x} = \frac{\partial w(l,t)}{\partial x} = 0 \tag{22}$$

To obtain the governing differential equation of the motion in its non-dimensional form following non-dimensional parameters are used:

$$\hat{w} = \frac{w}{g_0} \; , \; \hat{x} = \frac{x}{l} \; , \; \hat{t} = \frac{t}{\tilde{t}} \; , \; \hat{\Omega} = \Omega \tilde{t} \tag{23}$$

where:

$$\tilde{t} = \sqrt{\frac{(\rho A)_{eq} l^4}{(EI)_{eq}}} \tag{24}$$

Substituting equation (23) in equation (21), and considering viscous damping effect due to the squeeze film damping [12] and dropping the hats the non-dimensional differential equation of the motion reduces to:

$$\frac{\partial^4 w(x,t)}{\partial x^4} + \frac{\partial^2 w(x,t)}{\partial t^2} - [\alpha_1 + \alpha_2 \Gamma(w,w)]\frac{\partial^2 w(x,t)}{\partial x^2} + \alpha_3 \frac{\partial w(x,t)}{\partial t} = \alpha_4 \left( \frac{[V_{DC} + V_{AC}\sin(\Omega t)]^2}{(1-w)^2} - \frac{V_{DC}^2}{(1+w)^2} \right) \quad (25)$$

Where the following non-dimensional boundary conditions holds:

$$w(0,t) = w(1,t) = 0 \; , \; \frac{\partial w(0,t)}{\partial x} = \frac{\partial w(1,t)}{\partial x} = 0 \quad (26)$$

For simplicity in equation(25) the function $\Gamma$ and the coefficients $\alpha_i$ are defined as:

$$\Gamma(f_1(x,t), f_2(x,t)) = \int_0^1 \frac{\partial f_1}{\partial x} \frac{\partial f_2}{\partial x} dx \quad (27)$$

$$\alpha_1 = \frac{F_p l^2}{(EI)_{eq}} \; , \; \alpha_2 = \frac{(EA)_{eq} g_0^2}{2(EI)_{eq}} \; , \; \alpha_3 = \frac{\hat{c} l^2}{\sqrt{(\rho A)_{eq}(EI)_{eq}}} \; , \; \alpha_4 = \frac{\varepsilon_0 a l^4}{2 g_0^3 (EI)_{eq}}$$

## 3. Numerical Solution

Both sides of equation **(25)** are multiplied in the denominator of the electrostatic force, which yields

$$(1-w)^2 (1+w)^2 \frac{\partial^4 w}{\partial x^4} + (1-w)^2 (1+w)^2 \frac{\partial^2 w}{\partial t^2} - (1-w)^2 (1+w)^2 [\alpha_1 + \alpha_2 \Gamma(w,w)]\frac{\partial^2 w}{\partial x^2} \quad (28)$$
$$+ (1-w)^2 (1+w)^2 \alpha_3 \frac{\partial w}{\partial t}$$
$$= \alpha_4 V_{DC}^2 \left[ (1+w)^2 - (1-w)^2 \right] + (1+w)^2 2\alpha_4 V_{DC} V_{AC} \sin(\Omega t) + \alpha_4 (1+w)^2 V_{AC}^2 \sin^2(\Omega t)$$

Equation (28) reduces to:

$$\left(1-2w^{2}+w^{4}\right)\frac{\partial^{4}w}{\partial x^{4}}+\left(1-2w^{2}+w^{4}\right)\frac{\partial^{2}w}{\partial t^{2}}-\left(1-2w^{2}+w^{4}\right)\left[\alpha_{1}+\alpha_{2}\Gamma(w,w)\right]\frac{\partial^{2}w}{\partial x^{2}} \quad (29)$$
$$+\left(1-2w^{2}+w^{4}\right)\alpha_{3}\frac{\partial w}{\partial t}$$
$$=\alpha_{4}V_{DC}^{2}\left(4w\right)+\left(1+2w+w^{2}\right)\left(2\alpha_{4}V_{DC}V_{AC}\sin(\Omega t)+\alpha_{4}V_{AC}^{2}\sin^{2}(\Omega t)\right)$$

Based on the Galerkin method the following approximate solution is substituted in equation (29).

$$w(x,t)=\sum_{m=1}^{M}q_{m}(t)\varphi_{m}(x) \quad (30)$$

where $q_{m}(t)$ is the $m$th generalized coordinate and $\varphi_{m}(x)$ is the $m$th normalized linear un-damped mode shape of the straight micro-beam.

Applying Galerkin technique and considering $\varphi_{m}^{IV}(x)=\alpha_{1}\varphi_{m}''+\omega_{m}^{2}\varphi_{m}$, where $\omega_{m}$ is the $m$th natural frequency of the straight micro-beam, and based on the orthogonality of the normalized mode shapes as $\int_{0}^{1}\varphi_{m}(x)\varphi_{n}(x)dx=\delta_{ij}$, the reduced order differential equation of the micro-beam reduces to:

$$\omega_n^2 q_n + \alpha_1 \sum_{m=1}^{M} q_m \int_0^1 \varphi_n \varphi_m'' dx - 2\alpha_1 \sum_{i=1}^{M}\sum_{j=1}^{M}\sum_{m=1}^{M} q_i q_j q_m \int_0^1 \varphi_n \varphi_i \varphi_j \varphi_m'' dx - 2\sum_{i=1}^{M}\sum_{j=1}^{M}\sum_{m=1}^{M} \omega_m^2 q_i q_j q_m \int_0^1 \varphi_n \varphi_i \varphi_j \varphi_m dx \quad (31)$$

$$+\alpha_1 \sum_{i=1}^{M}\sum_{j=1}^{M}\sum_{k=1}^{M}\sum_{l=1}^{M}\sum_{m=1}^{M} q_i q_j q_k q_l q_m \int_0^1 \varphi_n \varphi_i \varphi_j \varphi_k \varphi_l \varphi_m'' dx + \sum_{i=1}^{M}\sum_{j=1}^{M}\sum_{k=1}^{M}\sum_{l=1}^{M}\sum_{m=1}^{M} \omega_m^2 q_i q_j q_k q_l q_m \int_0^1 \varphi_n \varphi_i \varphi_j \varphi_k \varphi_l \varphi_m dx$$

$$+\ddot{q}_n - 2\sum_{i=1}^{M}\sum_{j=1}^{M}\sum_{m=1}^{M} q_i q_j \ddot{q}_m \int_0^1 \varphi_n \varphi_i \varphi_j \varphi_m dx + \sum_{i=1}^{M}\sum_{j=1}^{M}\sum_{k=1}^{M}\sum_{l=1}^{M}\sum_{m=1}^{M} q_i q_j q_k q_l \ddot{q}_m \int_0^1 \varphi_n \varphi_i \varphi_j \varphi_k \varphi_l \varphi_m dx$$

$$-\alpha_1 \sum_{m=1}^{M} q_m \int_0^1 \varphi_n \varphi_m'' dx + 2\alpha_1 \sum_{i=1}^{M}\sum_{j=1}^{M}\sum_{m=1}^{M} q_i q_j q_m \int_0^1 \varphi_n \varphi_i \varphi_j \varphi_m'' dx$$

$$-\alpha_1 \sum_{i=1}^{M}\sum_{j=1}^{M}\sum_{k=1}^{M}\sum_{l=1}^{M}\sum_{m=1}^{M} q_i q_j q_k q_l q_m \int_0^1 \varphi_n \varphi_i \varphi_j \varphi_k \varphi_l \varphi_m'' dx$$

$$-\alpha_2 \sum_{m=1}^{M}\sum_{p=1}^{M}\sum_{r=1}^{M} q_m q_p q_r \int_0^1 \varphi_n \varphi_m'' \int_0^1 \varphi_p' \varphi_r' dx dx + 2\alpha_2 \sum_{i=1}^{M}\sum_{j=1}^{M}\sum_{m=1}^{M}\sum_{p=1}^{M}\sum_{r=1}^{M} q_i q_j q_m q_p q_r \int_0^1 \varphi_n \varphi_m'' \varphi_i \varphi_j \int_0^1 \varphi_p' \varphi_r' dx dx$$

$$-\alpha_2 \sum_{i=1}^{M}\sum_{j=1}^{M}\sum_{k=1}^{M}\sum_{l=1}^{M}\sum_{m=1}^{M}\sum_{p=1}^{M}\sum_{r=1}^{M} q_i q_j q_k q_l q_m q_p q_r \int_0^1 \varphi_n \varphi_i \varphi_j \varphi_k \varphi_l \varphi_m'' \int_0^1 \varphi_p' \varphi_r' dx dx$$

$$+\alpha_3 \dot{q}_n - 2\alpha_3 \sum_{i=1}^{M}\sum_{j=1}^{M}\sum_{m=1}^{M} q_i q_j \dot{q}_m \int_0^1 \varphi_n \varphi_i \varphi_j \varphi_m dx + \alpha_3 \sum_{i=1}^{M}\sum_{j=1}^{M}\sum_{k=1}^{M}\sum_{l=1}^{M}\sum_{m=1}^{M} q_i q_j q_k q_l \dot{q}_m \int_0^1 \varphi_n \varphi_i \varphi_j \varphi_k \varphi_l \varphi_m dx$$

$$-2\alpha_4 \left[ \left( V_{DC} + V_{AC} \sin(\Omega t) \right)^2 + V_{DC}^2 \right] q_n$$

$$-\alpha_4 \left[ \left( V_{DC} + V_{AC} \sin(\Omega t) \right)^2 - V_{DC}^2 \right] \sum_{i=1}^{M}\sum_{j=1}^{M} q_i q_j \int_0^1 \varphi_n \varphi_i \varphi_j dx$$

$$= \alpha_4 \left[ \left( V_{DC} + V_{AC} \sin(\Omega t) \right)^2 - V_{DC}^2 \right] \int_0^1 \varphi_n dx$$

To find a period solution of equation (31), we have applied the shooting method [24, 25]. Once a periodic solution is captured, its stability is investigated by examining the eigenvalues of the monodromy matrix [25, 26].

As a case study we have studied a fully clamped FGP micro beam (as indicated in Figure 1). The frequency response curves depict that for low amplitudes of harmonic excitation in the vicinity of the primary resonance the system exhibits hardening behavior as the frequency is swept forward. This behavior is due to the hardening effect of the geometric nonlinearity (mid-plane stretching) which resembles a nonlinear cubic type of stiffness [26]. As mentioned two main sources of nonlinearity including geometric nonlinearity and non-linear electro-static force govern the dynamics of the micro-beam. The geometric non-linearity has a hardening effect while the electro-static force has a softening effect [8, 10].

Figure 2, represents the frequency response curves for the case of $P_{p_u} = 0.50$ and $V_{DC}=2v$ $V_{AC}=10mv$ and various piezoelectric actuations.

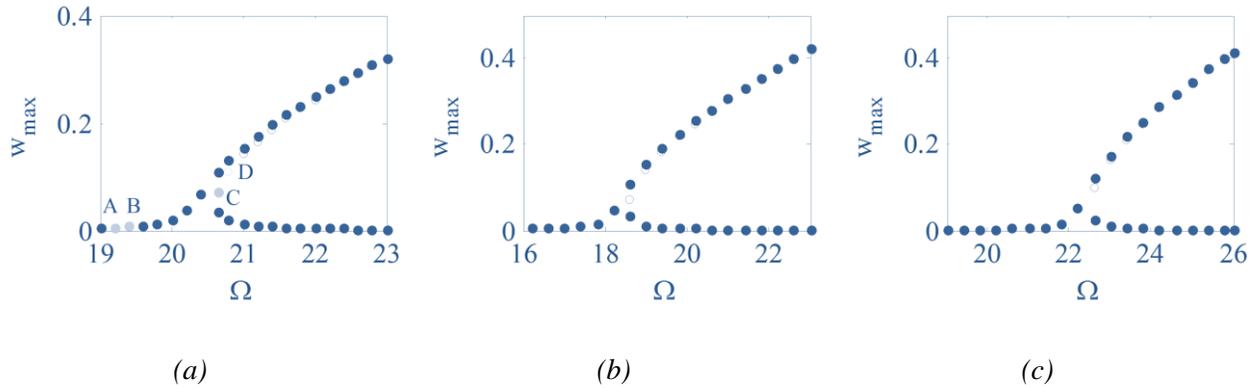

(a)          (b)          (c)

Figure 2: Frequency response curve representing the hardening effect near primary resonance (Filled circles represent the stable periodic solutions) $V_{DC}=2.0V$ $V_{AC}=10mV$, $P_{p_u} = 0.50$ (a): $V_P=0.0mV$, (b): $V_P=-90.0mV$, (c): $V_P=90.0mV$.

As the frequency response curves depict, based on the polarity of piezoelectric actuation, the frequency response curves can be tuned both in forward and backward directions. Piezoelectric actuation with negative polarity imposes a compressive axial force along the micro-beam and accordingly the frequency response curve moves backward in compare with $V_P=0.0mv$. Figure 3, illustrates the frequency response curves for the same $V_{DC}$, and $P_{p_u}$ as of Figure 2, and $V_{AC}=200mv$, and various piezoelectric voltages.

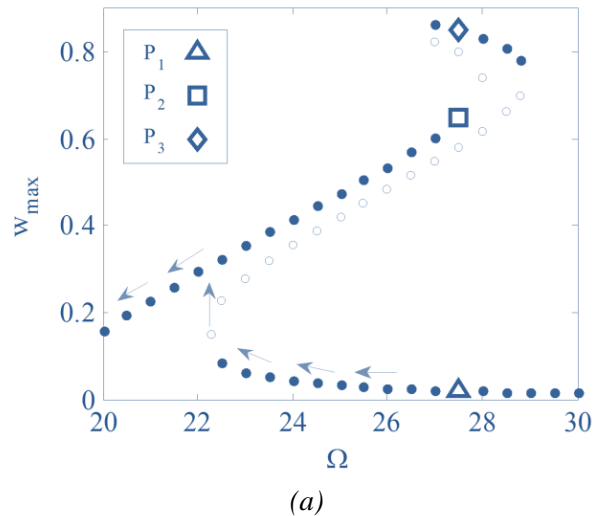

(a)

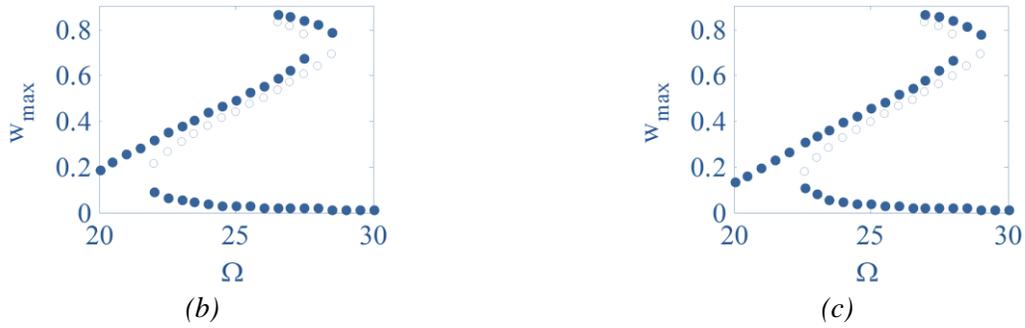

(b)  (c)

Figure 3: Frequency response curve representing the hardening effect near primary resonance (Filled circles represent the stable periodic solutions) $V_{DC}$=2.0V, $V_{AC}$=200mV, $P_{P_u}=0.50$ *(a):* $V_P$=0.0mV, *(b):* $V_P$=-20.0mV, *(c):* $V_P$=20.0mV.

Comparing the frequency response curves depicted in Figure 2 and Figure 3 reveal that both softening and hardening effects appear for higher $V_{AC}$s whereas for lower amplitudes of harmonic excitation the micro-resonator exhibits hardening effect, this qualitative behavior is also reported in the literature [8]. Furthermore, for lower amplitudes of harmonic excitation there is one cyclic fold bifurcation, however higher amplitudes lead to three cyclic fold bifurcation points on the frequency response curves.

In nonlinear systems corresponding to a determined excitation frequency, the system may have multi attractors (known as stable periodic solutions or stable limit cycles), assuming a bounded response condition, the steady state response and the amplitude of the steady state response depends on the fact that the applied initial condition is located on the basin of which attractor. Figure 4, depicts the basins of attractions for the three stable periodic solutions (attractors) corresponding to $\Omega=27.5$ which are illustrated in Figure 3-(b).

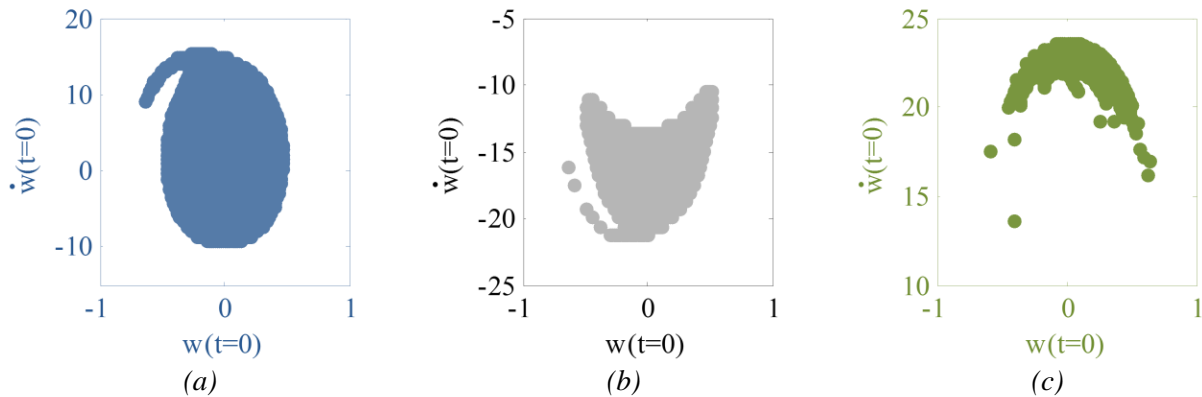

(a)  (b)  (c)

Figure 4: The basins of the periodic attractors corresponding to $V_{DC}$=2.0V, $V_{AC}$=200mV, $V_P$=0.0mV, $P_{P_u}=0.50$, $\Omega=27.5$, *(a):* $P_1$ *(b):* $P_2$, *(c):* $P_3$.

The basin of attraction becomes smaller as the amplitude of the periodic attractor increases.

## 4. Conclusion

The study focused on the nonlinear dynamics of a Functionally Graded Piezoelectric micro-resonator subject to a two side electrostatic and a simultaneous piezoelectric actuation as a micro-resonator. The shooting method was applied to detect the periodic. The dynamics of the micro-resonator in the vicinity of the primary resonance were studied; for low enough amplitudes of the harmonic excitation, the effect of geometric nonlinearity were dominant and as a result the frequency response curves were of hardening type; regardless of the polarity of the piezoelectric voltage we saw one cyclic-fold bifurcation in the frequency response curves where the Floquet exponents laid on the unit circle and the periodic orbits became of non-hyperbolic type. The results of the present study can be used as a design tool of tunable micro-resonators.